\documentclass[a4paper,11pt]{article}
\usepackage{pos}
\usepackage{amsfonts}
\usepackage{latexsym}
\usepackage{verbatim}
\usepackage{amsthm}
\usepackage{amsmath}
\usepackage{wrapft}
\usepackage{mathrsfs}
\newcommand{\NF}{{{\mathbb N}}}
\newcommand{\RF}{{{\mathbb R}}}
\newcommand{\TF}{{{\mathbb T}}}
\newcommand{\ScrF}{{{\mathscr F}}}
\newcommand{\ScrG}{{{\mathscr G}}}
\newcommand{\ScrM}{{{\mathscr M}}}
\newcommand{\ScrQ}{{{\mathscr Q}}}
\newcommand{\ScrT}{{{\mathscr T}}}
\newcommand{\ScrX}{{{\mathscr X}}}
\newcommand{\ScrY}{{{\mathscr Y}}}
\newtheorem{proposal}{Proposal}%[section]
\newtheorem{conjecture}{Conjecture}%[section]
\title{ICRC2023 Proceedings : Proposal of a gauge-invariant treatment
  of $l=0,1$-mode perturbations on the Schwarzschild background spacetime}
 \ShortTitle{Proposal of a gauge-invariant treatment of $l=0,1$-mode perturbations on ...}

\author*[a]{Kouji Nakamura}
\affiliation[a]{Gravitational-Wave Science Project, National
  Astronomical Observatory of Japan, \\
  2-21-1, Osawa, Mitaka, Tokyo 181-8588, Japan}

\emailAdd{dr.kouji.nakamura@gmail.com}
\abstract{
  A gauge-invariant perturbation theory on a generic background
  spacetime is developing from 2003 and ``zero-mode problem'' for
  linear metric perturbations was proposed as the essential problem of
  this theory.
  In the perturbation theory on the Schwarzschild background
  spacetime, $l=0,1$ modes correspond to the above ``zero-mode'' and
  the gauge-invariant treatments of these modes is a famous
  non-trivial problem in perturbation theories on the Schwarzschild
  background spacetime.
  Due to this situation, a gauge-invariant treatment for these
  $l=0,1$-mode perturbations is proposed.
  Through this gauge-invariant treatment, the solutions to the
  linearized Einstein equation for these modes with a generic matter
  field are derived.
  In the vacuum case, the linearized version of uniqueness theorem of
  Kerr spacetime is confirmed in a gauge-invariant manner.
  In this sense, our proposal is reasonable.
}

\ConferenceLogo{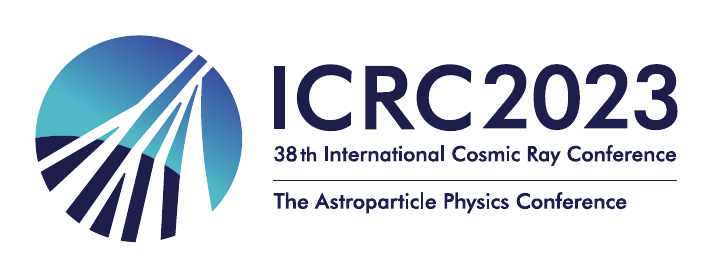}

\FullConference{%
38th International Cosmic Ray Conference (ICRC2023)\\
  26 July - 3 August, 2023\\
  Nagoya, Japan}

\begin{document}
\maketitle

%%%%%%%%%%%%%%%%%%%%%%%%%%%%%%%%%%%%%%%%%%%%%%%%%%%%%%
%%%%%%%%%%%%%%%%%%%%%%%%%%%%%%%%%%%%%%%%%%%%%%%%%%%%%%
%%%%%%%%%%%%%%%%%%%%%%%%%%%%%%%%%%%%%%%%%%%%%%%%%%%%%%
%%%%%%%%%%%%%%%%%%%%%%%%%%%%%%%%%%%%%%%%%%%%%%%%%%%%%%
%\section{Introduction}
%\label{sec:introduction}
%%%%%%%%%%%%%%%%%%%%%%%%%%%%%%%%%%%%%%%%%%%%%%%%%%%%%%
%%%%%%%%%%%%%%%%%%%%%%%%%%%%%%%%%%%%%%%%%%%%%%%%%%%%%%
%%%%%%%%%%%%%%%%%%%%%%%%%%%%%%%%%%%%%%%%%%%%%%%%%%%%%%
%%%%%%%%%%%%%%%%%%%%%%%%%%%%%%%%%%%%%%%%%%%%%%%%%%%%%%

%*************************************************************

\paragraph{\bf 1. Introduction} ----------
From the direct observation of gravitational
waves~\cite{LIGO-GW150914-2016}, in 2015, the gravitational-wave
astronomy and multi-messenger astronomy including gravitational waves
began.
One of future directions of gravitational-wave astronomy
is the development as a precise science by the detailed studies of
source science and the tests of general-relativity.
To support such precise sciences, higher-order perturbation theories
in general relativity are useful.

%*************************************************************

Among future targets of gravitational-wave sources, the
Extreme-Mass-Ratio-Inspiral (EMRI) is one of the targets of the Laser
Interferometer Space Antenna~\cite{LISA-homepage}.
The EMRI is a source of gravitational waves, which is the motion
of a stellar mass object around a supermassive black hole, and black
hole perturbation theories are used to describe this EMRI.
Therefore, theoretical sophistications of black hole perturbation
theories and their higher-order extensions are necessary.

%*************************************************************

Although realistic black holes have their angular momentum and we must
consider the perturbation theory of a Kerr black hole for direct
applications to the EMRI, further sophistication is possible even in
perturbation theories on the Schwarzschild spacetime.
Based on the pioneering works by Regge and
Wheeler, and Zerilli~\cite{T.Regge-J.A.Wheeler-1957}, there have been
many studies on the perturbations of the Schwarzschild spacetime.
Because the Schwarzschild spacetime has the spherical symmetry, we
decompose perturbations through the spherical harmonics $Y_{lm}$ and
classify them into odd- and even-modes based on their parity.
However, $l=0$ and $l=1$ modes should be separately treated, and
``{\it gauge-invariant}'' treatments for $l=0$ and $l=1$ even-modes
remain unknown.

%*************************************************************

In this situation, we proposed a gauge-invariant treatment of
$l=0,1$-modes and derived the solutions to the linearized Einstein
equations for these modes~\cite{K.Nakamura-2021a}.
The obtained solutions~\cite{K.Nakamura-2021a} are physically
reasonable.
For this reason, we may say that our proposal is also reasonable.
In addition, owing to our proposal, the formulation of higher-order
gauge-invariant perturbation theory developed
in~\cite{K.Nakamura-2003,K.Nakamura-2011,K.Nakamura-2014}
becomes applicable to any-order perturbations on the Schwarzschild
background spacetime~\cite{K.Nakamura-2021b}.
In this manuscript, we briefly explain these issues.

%*************************************************************

%%%%%%%%%%%%%%%%%%%%%%%%%%%%%%%%%%%%%%%%%%%%%%%%%%%%%%
%%%%%%%%%%%%%%%%%%%%%%%%%%%%%%%%%%%%%%%%%%%%%%%%%%%%%%
%%%%%%%%%%%%%%%%%%%%%%%%%%%%%%%%%%%%%%%%%%%%%%%%%%%%%%
%\section{Brief review of general-relativistic gauge-invariant perturbation theory}
%\label{sec:review-of-perturbation-theroy}
%%%%%%%%%%%%%%%%%%%%%%%%%%%%%%%%%%%%%%%%%%%%%%%%%%%%%%
%%%%%%%%%%%%%%%%%%%%%%%%%%%%%%%%%%%%%%%%%%%%%%%%%%%%%%
%%%%%%%%%%%%%%%%%%%%%%%%%%%%%%%%%%%%%%%%%%%%%%%%%%%%%%

%*********************************************************************

\paragraph{\bf 2. Brief review of general-relativistic gauge-invariant perturbation theory} ----------
General relativity is a theory based on general covariance, and that
covariance is the reason that the notion of ``gauge'' has been
introduced into the theory.
In particular, in general relativistic perturbations,
{\it the second-kind gauge} appears in
perturbations~\cite{K.Nakamura-2010}.
In general-relativistic perturbation theory, we usually treat the
one-parameter family of spacetimes
$\{(\ScrM_{\lambda},Q_{\lambda})|\lambda\in[0,1]\}$ to discuss
differences between the background spacetime
$(\ScrM,Q_{0})$ $=$ $(\ScrM_{\lambda=0},Q_{\lambda=0})$ and the
physical spacetime $(\ScrM_{{\rm ph}},\bar{Q})$ $=$
$(\ScrM_{\lambda=1},Q_{\lambda=1})$.
Here, $\lambda$ is the infinitesimal parameter for perturbations,
$\ScrM_{\lambda}$ is a spacetime manifold for each $\lambda$, and
$Q_{\lambda}$ is the collection of the tensor fields on
$\ScrM_{\lambda}$.
Since each $\ScrM_{\lambda}$ is a different manifold, we have to
introduce the point identification map $\ScrX_{\lambda}$ $:$
$\ScrM\rightarrow\ScrM_{\lambda}$ to compare tensor fields on
different manifolds.
This point-identification is
{\it the gauge choice of the second kind}.
Since we have no guiding principle by which to choose identification
map $\ScrX_{\lambda}$ due to the general covariance, we may choose
a different point-identification $\ScrY_{\lambda}$ from
$\ScrX_{\lambda}$.
This degree of freedom in the gauge choice is {\it the gauge degree of
  freedom of the second kind.}
{\it The gauge-transformation of the second kind} is a change of this
identification map.
We note that this second-kind gauge is a different notion of the
degree of freedom of coordinate choices on a single manifold, which is
called {\it the gauge of the first
  kind}~\cite{K.Nakamura-2010}.

%*********************************************************************

Once we introduce the second-kind gauge choice $\ScrX_{k}$ $:$
$\ScrM$ $\rightarrow$ $\ScrM_{\lambda}$, we can compare the
tensor fields on different manifolds $\{\ScrM_{\lambda}\}$, and
{\it perturbations} of a tensor field $Q_{\lambda}$ are represented by
the difference $\ScrX_{\lambda}^{*}Q_{\lambda} - Q_{0}$, where
$\ScrX_{\lambda}^{*}$ is the pull-back induced by the gauge choice
$\ScrX_{\lambda}$ and $Q_{0}$ is the background value of the
variable $Q_{\lambda}$.
This representation of perturbations completely depends on
$\ScrX_{\lambda}$.
If we change the gauge choice from $\ScrX_{\lambda}$ to ${\cal
  Y}_{\lambda}$, the pulled-back variable of $Q_{\lambda}$ is
represented by $\ScrY_{\lambda}^{*}Q_{\lambda}$.
These different representations are related through the
gauge-transformation rule
\begin{eqnarray}
  \label{eq:gauge-trans-from-calYQ-to-calXQ}
  \ScrY_{\lambda}^{*}Q_{\lambda}
  =
  \Phi^{*}_{\lambda} \ScrX_{\lambda}^{*}Q_{\lambda}
  ,
  \quad
  \Phi_{\lambda} := \ScrX_{\lambda}^{-1} \circ \ScrY_{\lambda}.
\end{eqnarray}
$\Phi_{\lambda}$ is a diffeomorphism on the background spacetime
$\ScrM$.

%*********************************************************************

In the perturbative approach, we treat the perturbations of the
pulled-back variable $\ScrX_{\lambda}^{*}Q_{\lambda}$ through the
Taylor series with respect to the infinitesimal parameter $\lambda$ as
\begin{eqnarray}
  \label{eq:calXQ-expand}
  \ScrX_{\lambda}^{*}Q_{\lambda}
  =:
  \sum_{n=0}^{k} \frac{\lambda^{n}}{n!} {}^{(n)}_\ScrX\!Q
  +
  O(\lambda^{k+1}),
\end{eqnarray}
where ${}^{(n)}_{\ScrX}\!Q$ is the representation of the $k$th-order
perturbation of $Q_{\lambda}$ under $\ScrX_{\lambda}$ with
${}^{(0)}_{\ScrX}\!Q=Q_{0}$.

%*********************************************************************

Similarly, we can have the representation of the perturbation of
$Q_{\lambda}$ under the different gauge choice $\ScrY_{\lambda}$
from $\ScrX_{\lambda}$.
Since these different representations are related to the
gauge-transformation rule (\ref{eq:gauge-trans-from-calYQ-to-calXQ}),
the order-by-order gauge-transformation rule between
${}^{(n)}_\ScrX\!Q$ and ${}^{(n)}_\ScrY\!Q$ is
given from the Taylor expansion of
Eq.~(\ref{eq:gauge-trans-from-calYQ-to-calXQ}).
In general, $\Phi_{\lambda}$ is given by a {\it knight
  diffeomorphism}~\cite{K.Nakamura-2003}:
{\it Let $\Phi_{\lambda}$ be a one-parameter family of
  diffeomorphisms, and $T$ a tensor field such that
  $\Phi_{\lambda}^{*}T$ is of class $C^{k}$.
  Then, $\Phi_{\lambda}^{*}T$ can be expanded around $\lambda=0$ as
  \begin{eqnarray}
    \label{eq:PhilambdaastT-general-expansion}
    \Phi_{\lambda}^{*}T
    =
    \sum_{n=0}^{k} \lambda^{n}
    \sum_{\{j_{i}\}\in J_{n}} C_{n,\{j_{i}\}}
    {\pounds}_{\xi_{(1)}}^{j_{1}}\cdots{\pounds}_{\xi_{(n)}}^{j_{n}}T
    +
    O(\lambda^{k+1})
    .
  \end{eqnarray}
  Here, $J_{n}:=\left\{ \{j_{i}\} | {}^{\forall}i \in \NF, j_{i}\in \NF,
    s.t. \sum_{i=1}^{\infty} ij_{i}=n\right\}$ and
  $C_{n,\{j_{i}\}}$ $:=$ $\displaystyle \prod_{i=1}^{n}
  \frac{1}{(i!)^{j_{i}}j_{i}!}$.
  The vector fields $\xi_{(1)}$, $...$, $\xi_{(k)}$ in
  Eq.~(\ref{eq:PhilambdaastT-general-expansion}) are called the
  generators of $\Phi_{\lambda}$.
}

%*********************************************************************

Substituting Eqs.~(\ref{eq:calXQ-expand}) and
(\ref{eq:PhilambdaastT-general-expansion}) into
Eq.~(\ref{eq:gauge-trans-from-calYQ-to-calXQ}), we obtain the
order-by-order gauge-transformation rules between
${}^{(n)}_{\ScrX}\!Q$ and ${}^{(n)}_{\ScrY}\!Q$ as
\begin{eqnarray}
  \label{eq:nth-order-gauge-trans}
  {}^{(n)}_{\;\;\ScrY}\!Q - {}^{(n)}_{\;\;\ScrX}\!Q
  =
  \sum_{l=1}^{n} \frac{n!}{(n-l)!} \sum_{\{j_{i}\}\in J_{l}}
  C_{l,\{J_{i}\}}
  {\pounds}_{\xi_{(1)}}^{j_{1}}
  \cdots
  {\pounds}_{\xi_{(l)}}^{j_{l}}
  {}^{(n-l)}_{\;\;\;\;\;\ScrX}\!Q
  .
\end{eqnarray}

%*********************************************************************

Inspecting the gauge-transformation rule
(\ref{eq:nth-order-gauge-trans}), we first defined gauge-invariant
variables for metric
perturbations~\cite{K.Nakamura-2003}.
We consider the metric $\bar{g}_{ab}$ on
$(\ScrM_{{\rm ph}},\bar{Q})$ $=$
$(\ScrM_{\lambda=1},Q_{\lambda=1})$,
and we expand the pulled-back metric
$\ScrX_{\lambda}^{*}\bar{g}_{ab}$ to $\ScrM$ through a gauge
choice $\ScrX_{k}$ as
\begin{eqnarray}
  \label{eq:full-metric-expansion}
  \ScrX_{\lambda}\bar{g}_{ab}
  =
  \sum_{n=0}^{k} \frac{\lambda^{n}}{n!} {}^{(n)}_{\ScrX}g_{ab} + O(\lambda^{k+1}),
\end{eqnarray}
where $g_{ab}:={}^{(0)}_{\ScrX}g_{ab}$ is the metric on $\ScrM$.
The expansion (\ref{eq:full-metric-expansion}) of the metric depends
entirely on $\ScrX_{\lambda}$.
Nevertheless, henceforth, we do not explicitly express the index of
the gauge choice $\ScrX_{\lambda}$ if there is no possibility of
confusion.
In~\cite{K.Nakamura-2003}, we proposed a
procedure to construct gauge-invariant variables for higher-order
perturbations.
Our starting point of this construction was the following conjecture
for the linear metric perturbation $h_{ab}:={}^{(1)}\!g_{ab}$:
\begin{conjecture}
  \label{conjecture:decomposition-conjecture}
  If the gauge-transformation rule for a pulled-back tensor field
  $h_{ab}$ from $\ScrM_{\rm ph}$ to $\ScrM$ is given by
  ${}_{\ScrY}\!h_{ab}$ $-$ ${}_{\ScrX}\!h_{ab}$ $=$
  ${\pounds}_{\xi_{(1)}}g_{ab}$ with the metric $g_{ab}$ on
  $\ScrM$, there then exist a tensor field $\ScrF_{ab}$ and a
  vector field $Y^{a}$ such that $h_{ab}$ is given by $h_{ab}$
  $=:$ $\ScrF_{ab}$ $+$ ${\pounds}_{Y}g_{ab}$, where
  $\ScrF_{ab}$ and $Y^{a}$ are transformed as
  ${}_{\ScrY}\!\ScrF_{ab}$ $-$ ${}_{\ScrX}\!\ScrF_{ab}$
  $=$ $0$ and ${}_{\ScrY}\!Y^{a}$ $-$ ${}_{\ScrX}\!Y^{a}$ $=$
  $\xi^{a}_{(1)}$ under the gauge transformation, respectively.
\end{conjecture}
We call $\ScrF_{ab}$ and $Y^{a}$ as the
{\it gauge-invariant} and {\it gauge-variant} parts
of $h_{ab}$, respectively.

%*********************************************************************

Based on Conjecture~\ref{conjecture:decomposition-conjecture},
in~\cite{K.Nakamura-2014}, we found that the $n$th-order metric
perturbation ${}^{(n)}_{\ScrX}g_{ab}$ is decomposed into its
gauge-invariant and gauge-variant parts
as~\footnote{
  Precisely speaking, to reach to the decomposition formula
  (\ref{eq:nth-order-original-ngab-decomp}), we have to confirm
  Conjecture 4.1 in Ref.~\cite{K.Nakamura-2014} in addition to
  Conjecture~\ref{conjecture:decomposition-conjecture}.
}
\begin{eqnarray}
  {}^{(n)}\!g_{ab}
  &=&
    {}^{(n)}\!\ScrF_{ab}
     -
     \sum_{l=1}^{n}
     \frac{n!}{(n-l)!}
     \sum_{\{j_{i}\}\in J_{l}}
     C_{l,\{j_{i}\}}
     {\pounds}_{-{}^{(1)}\!Y}^{j_{1}}\cdots{\pounds}_{-{}^{(l)}\!Y}^{j_{l}}
     {}^{(n-l)}\!g_{ab}
     .
     \label{eq:nth-order-original-ngab-decomp}
\end{eqnarray}
Furthermore, through the gauge-variant variables ${}^{(i)}Y^{a}$
($i=1,...,n$), we also found the definition of the gauge-invariant
variable ${}^{(n)}\!\ScrQ$ for the $n$th-order perturbation
${}^{(n)}\!Q$ of an arbitrary tensor field $Q$.
This definition of the gauge-invariant variable ${}^{(n)}\!\ScrQ$
implies that the $n$th-order perturbation ${}^{(n)}\!Q$ of any tensor
field $Q$ is always decomposed into its gauge-invariant part and
gauge-variant part as
\begin{eqnarray}
  {}^{(n)}\!Q
  =
  {}^{(n)}\!\ScrQ
  -
  \sum_{l=1}^{n}
  \frac{n!}{(n-l)!}
  \sum_{\{j_{i}\}\in J_{l}}
  C_{l,\{j_{i}\}}
  {\pounds}_{-{}^{(1)}\!Y}^{j_{1}}\cdots{\pounds}_{-{}^{(l)}\!Y}^{j_{l}}
  {}^{(n-l)}\!Q
  .
  \label{eq:nth-order-original-nQ-decomp}
\end{eqnarray}

%*********************************************************************

For example, the perturbative expansion of the Einstein tensor and
the energy-momentum tensor, which are pulled back through
$\ScrX_{\lambda}$, are given by
\begin{eqnarray}
  \label{eq:barGab-expansion}
  \ScrX_{\lambda}^{*}\bar{G}_{a}^{\;\;b}
  =
      \sum_{n=0}^{k} \frac{\lambda^{n}}{n!}
      {}^{(n)}_{\ScrX}\!G_{a}^{\;\;b}
      +
      O(\lambda^{k+1})
      ,
  \quad
  % \label{eq:arTab-expansion}
  \ScrX_{\lambda}^{*}\bar{T}_{a}^{\;\;b}
  =
      \sum_{n=0}^{k} \frac{\lambda^{n}}{n!}
      {}^{(n)}_{\ScrX}\!T_{a}^{\;\;b}
      +
      O(\lambda^{k+1})
      .
\end{eqnarray}
Then, the $n$th-order perturbation
${}^{(n)}_{\ScrX}G_{a}^{\;\;b}$ of the Einstein tensor and the
$n$th-order perturbation ${}^{(n)}_{\ScrX}T_{a}^{\;\;b}$ of the
energy-momentum tensor are also decomposed as
\begin{eqnarray}
  {}^{(n)}\!G_{a}^{\;\;b}
  &=&
    {}^{(n)}\!\ScrG_{a}^{\;\;b}
     -
     \sum_{l=1}^{n}
     \frac{n!}{(n-l)!}
     \sum_{\{j_{i}\}\in J_{l}}
     C_{l,\{j_{i}\}}
     {\pounds}_{-{}^{(1)}\!Y}^{j_{1}}\cdots{\pounds}_{-{}^{(l)}\!Y}^{j_{l}}
     {}^{(n-l)}\!G_{a}^{\;\;b}
     ,
     \label{eq:nth-order-original-nGab-decomp}
  \\
  {}^{(n)}\!T_{a}^{\;\;b}
  &=&
    {}^{(n)}\!\ScrT_{a}^{\;\;b}
     -
     \sum_{l=1}^{n}
     \frac{n!}{(n-l)!}
     \sum_{\{j_{i}\}\in J_{l}}
     C_{l,\{j_{i}\}}
     {\pounds}_{-{}^{(1)}\!Y}^{j_{1}}\cdots{\pounds}_{-{}^{(l)}\!Y}^{j_{l}}
     {}^{(n-l)}\!T_{a}^{\;\;b}
     .
     \label{eq:nth-order-original-nTab-decomp}
\end{eqnarray}
Through the lower-order Einstein equation
${}^{(k)}_{\ScrX}\!G_{a}^{\;\;b}=8\pi{}^{(k)}_{\ScrX}\!T_{a}^{\;\;b}$\footnote{
  We use the unit $G=c=1$, where $G$ is Newton's constant of
  gravitation, and $c$ is the velocity of light.
}
with $k\leq n-1$, the $n$th-order Einstein equation
${}^{(n)}_{\ScrX}\!G_{a}^{\;\;b}=8\pi{}^{(n)}_{\ScrX}\!T_{a}^{\;\;b}$
is automatically given in the gauge-invariant form
\begin{eqnarray}
  {}^{(n)}\!\ScrG_{a}^{\;\;b}
  =
  {}^{(1)}\!\ScrG_{a}^{\;\;b}\left[{}^{(n)}\!\ScrF\right]
  +
  {}^{({\rm NL})}\!\ScrG_{a}^{\;\;b}\left[\left\{\left.{}^{(i)}\!\ScrF\right|i<n\right\}\right]
  =
  8\pi {}^{(n)}\!\ScrT_{a}^{\;\;b},
  \label{eq:nth-order-Einstein-eq}
\end{eqnarray}
where ${}^{(1)}\!\ScrG_{a}^{\;\;b}$ is the gauge-invariant part of
the linear-order perturbation of the Einstein tensor.
Explicitly, ${}^{(1)}\!\ScrG_{a}^{\;\;b}\left[A\right]$ for an
arbitrary tensor field $A_{ab}$ of the second rank is given
by~\cite{K.Nakamura-2003}
\begin{eqnarray}
  \label{eq:linear-Einstein-AIA2010-2}
  &&
     {}^{(1)}\ScrG_{a}^{\;\;b}\left[A\right]
     :=
     {}^{(1)}\Sigma_{a}^{\;\;b}\left[A\right]
     - \frac{1}{2} \delta_{a}^{\;\;b} {}^{(1)}\Sigma_{c}^{\;\;c}\left[A\right]
     ,
  \\
  \label{eq:(1)Sigma-def-linear}
  &&
     {}^{(1)}\Sigma_{a}^{\;\;b}\left[A\right]
     :=
     - 2 \nabla_{[a}^{}H_{d]}^{\;\;\;bd}\left[A\right]
     - A^{cb} R_{ac}
     ,
  \quad
%  \label{eq:Habc-def-linear}
     H_{ba}^{\;\;\;\;c}\left[A\right]
     :=
     \nabla_{(a}A_{b)}^{\;\;\;\;c} - \frac{1}{2} \nabla^{c}A_{ab}
     .
\end{eqnarray}
As derived in~\cite{K.Nakamura-2003}, when the background Einstein
tensor vanishes, we obtain the identity
$\nabla_{a}{}^{(1)}\!\ScrG_{b}^{\;\;a}\left[A\right]$ $=$ $0$
for an arbitrary tensor field $A_{ab}$ of the second rank.

%*********************************************************************

We emphasize that Conjecture~\ref{conjecture:decomposition-conjecture}
was the important premise of the above framework of the higher-order
perturbation theory.

%*********************************************************************

%%%%%%%%%%%%%%%%%%%%%%%%%%%%%%%%%%%%%%%%%%%%%%%%%%%%%%
%%%%%%%%%%%%%%%%%%%%%%%%%%%%%%%%%%%%%%%%%%%%%%%%%%%%%%
%%%%%%%%%%%%%%%%%%%%%%%%%%%%%%%%%%%%%%%%%%%%%%%%%%%%%%
%\section{Linear perturbations on spherically symmetric background}
%\label{sec:spherical_background_case}
%%%%%%%%%%%%%%%%%%%%%%%%%%%%%%%%%%%%%%%%%%%%%%%%%%%%%%
%%%%%%%%%%%%%%%%%%%%%%%%%%%%%%%%%%%%%%%%%%%%%%%%%%%%%%
%%%%%%%%%%%%%%%%%%%%%%%%%%%%%%%%%%%%%%%%%%%%%%%%%%%%%%

%*********************************************************************

\paragraph{\bf 3. Linear perturbations on spherically symmetric background} ----------
We use the 2+2 formulation of the perturbations on spherically
symmetric spacetimes.
The topological space of spherically symmetric spacetimes is
$\ScrM=\ScrM_{1}\times S^{2}$, and the metric on
this spacetime is
\begin{eqnarray}
  g_{ab}
  =
  y_{ab} + r^{2}\gamma_{ab}
  ,
  \quad
  y_{ab} = y_{AB} (dx^{A})_{a}(dx^{B})_{b}
  ,
  \quad
  \gamma_{ab} = \gamma_{pq} (dx^{p})_{a} (dx^{q})_{b}
  ,
  % \label{eq:background-metric-2+2-separate}
  \label{eq:background-metric-2+2}
\end{eqnarray}
where $x^{A} = (t,r)$, $x^{p}=(\theta,\phi)$, and $\gamma_{pq}$ is a
metric of the unit sphere.
In the Schwarzschild spacetime,
$y_{ab}=-f(dt)_{a}(dt)_{b}+f^{-1}(dr)_{a}(dr)_{b}$ with $f=1-2M/r$.

%*******************************************************************

On this $(\ScrM,g_{ab})$, we consider the components of the metric
perturbation as
\begin{eqnarray}
%  &&
     h_{ab}
     =
     h_{AB} (dx^{A})_{a}(dx^{B})_{b}
     +
     2 h_{Ap} (dx^{A})_{(a}(dx^{p})_{b)}
%     \nonumber\\
%  && \quad\quad\quad
     +
     h_{pq} (dx^{p})_{a}(dx^{q})_{b}
     .
\end{eqnarray}
In Ref.~\cite{K.Nakamura-2021a}, we proposed the decomposition of
these components as
\begin{eqnarray}
  \label{eq:hAB-fourier}
  &&
     h_{AB}
     =
     \sum_{l,m} \tilde{h}_{AB} S_{\delta}
     ,
  \quad
  % &&
  % \label{eq:hAp-fourier}
     h_{Ap}
     =
     r \sum_{l,m} \left[
     \tilde{h}_{(e1)A} \hat{D}_{p}S_{\delta}
     +
     \tilde{h}_{(o1)A} \epsilon_{pq} \hat{D}^{q}S_{\delta}
     \right]
     ,
  \\
  &&
     \label{eq:hpq-fourier}
     h_{pq}
     =
     r^{2} \sum_{l,m} \left[
     \frac{1}{2} \gamma_{pq} \tilde{h}_{(e0)} S_{\delta}
     +
     \tilde{h}_{(e2)} \left(
     \hat{D}_{p}\hat{D}_{q} - \frac{1}{2} \gamma_{pq} \hat{\Delta}
     \right) S_{\delta}
%     \right.
%     \nonumber\\
%  && \quad\quad\quad\quad\quad\quad
%     \left.
     +
     2 \tilde{h}_{(o2)} \epsilon_{r(p} \hat{D}_{q)}\hat{D}^{r} S_{\delta}
     \right]
     ,
\end{eqnarray}
where $\hat{D}_{p}$ is the covariant derivative associated with
the metric $\gamma_{pq}$ on $S^{2}$,
$\hat{D}^{p}:=\gamma^{pq}\hat{D}_{q}$, and
$\epsilon_{pq}=\epsilon_{[pq]}$ is the totally antisymmetric
tensor on $S^{2}$.

%*************************************************************

The decomposition
(\ref{eq:hAB-fourier})--(\ref{eq:hpq-fourier}) implicitly state that
the Green functions of the
derivative operators $\hat{\Delta}:=\hat{D}^{r}\hat{D}_{r}$ and
$\hat{\Delta}+2:=\hat{D}^{r}\hat{D}_{r}+2$ should exist if we require
the one-to-one correspondence between $\{h_{Ap},$ $h_{pq}\}$ and
$\{\tilde{h}_{(e1)A},$ $\tilde{h}_{(o1)A},$ $\tilde{h}_{(e0)},$
$\tilde{h}_{(e2)},$ $\tilde{h}_{(o2)}\}$.
Because the eigenvalue of the operator $\hat{\Delta}$ on $S^{2}$ is
$-l(l+1)$, the kernels of the operators $\hat{\Delta}$ and
$\hat{\Delta}+2$ are $l = 0$ and $l = 1$ modes, respectively.
Thus, the one-to-one correspondence between
$\{h_{Ap},$ $h_{pq}\}$ and
$\{\tilde{h}_{(e1)A},$ $\tilde{h}_{(o1)A},$ $\tilde{h}_{(e0)},$
$\tilde{h}_{(e2)},$ $\tilde{h}_{(o2)}\}$ is not guaranteed for $l =
0,1$ modes in Eqs.~(\ref{eq:hAB-fourier})--(\ref{eq:hpq-fourier}) with
$S_{\delta}=Y_{lm}$.
To recover this one-to-one correspondence, we consider the
scalar harmonics~\cite{K.Nakamura-2021a}
\begin{eqnarray}
  \label{eq:extended-harmonic-functions}
  S_{\delta}
  =
  \left\{
  Y_{lm} \;\; \mbox{for} \;\; l\geq 2; \quad
  k_{(\hat{\Delta}+2)m} \;\; \mbox{for} \;\;  l=1; \quad
  k_{(\hat{\Delta})} \;\; \mbox{for} \;\; l=0
  \right\}
  .
\end{eqnarray}
As the explicit functions of $k_{(\hat{\Delta})}$ and
$k_{(\hat{\Delta}+2)m}$, we employ
\begin{eqnarray}
  \label{eq:l=0-general-mode-func-specific}
  &&
     k_{(\hat{\Delta})}
     =
     1
     +
     \delta \ln\left(\frac{1-z}{1+z}\right)^{1/2}
%     ,
%     \quad \delta\in\RF
     ,
     \quad
%  \\
%  \label{eq:l=1-m=0-general-mode-func-specific}
%  &&
     k_{(\hat{\Delta}+2)m=0}
     =
     z
     +
     \delta
     \left(\frac{z}{2}\ln\frac{1+z}{1-z}-1\right)
       ,
%     \quad
  \\
  \label{eq:l=1-m=pm1-general-mode-func-specific}
  &&
     k_{(\hat{\Delta}+2)m=\pm 1}
     =
     (1-z^{2})^{1/2}
     \left\{
     1
     +
     \delta
     \left(\frac{1}{2} \ln\frac{1+z}{1-z}+\frac{z}{1-z^{2}}\right)
     \right\} e^{\pm i \phi}
     ,
\end{eqnarray}
where $\delta\in\RF$ and $z = \cos\theta$.
This choice guarantees the linear-independence of the set
\begin{eqnarray}
  \label{eq:set-of-harmonic-functions}
  \left\{S_{\delta}, \hat{D}_{p}S_{\delta},
  \epsilon_{pq}\hat{D}^{q}S_{\delta},
  \displaystyle \frac{1}{2} \gamma_{pq}S_{\delta},
  \left(\hat{D}_{p}\hat{D}_{q}-\frac{1}{2}\gamma_{pq}\hat{\Delta}\right)S_{\delta},
  2\epsilon_{r(p}\hat{D}_{q)}\hat{D}^{r}S_{\delta}\right\}
\end{eqnarray}
of the harmonic functions including $l=0,1$ modes if $\delta\neq 0$,
but is singular if $\delta\neq 0$.
On the other hand, when $\delta = 0$, we have
$k_{(\hat{\Delta})}\propto Y_{00}$ and
$\hat{k}_{(\hat{\Delta}+2)m}\propto Y_{1m}$.

%*********************************************************************

Through the above harmonics functions $S_{\delta}$,
in Ref.~\cite{K.Nakamura-2021a}, we proposed the following strategy:
\begin{proposal}
  \label{proposal:harmonic-extension}
  We decompose the metric perturbations $h_{ab}$ on the background
  spacetime with the metric (\ref{eq:background-metric-2+2}),
  through Eqs.~(\ref{eq:hAB-fourier})--(\ref{eq:hpq-fourier}) with the
  harmonic functions $S_{\delta}$ given by
  Eq.~(\ref{eq:extended-harmonic-functions}).
  After deriving the mode-by-mode field equations such as linearized
  Einstein equations using $S_{\delta}$, we choose $\delta=0$ when we
  solve these field equations as the regularity of solutions.
\end{proposal}

%*********************************************************************

Once we accept
Proposal~\ref{proposal:harmonic-extension}, we can justify
Conjecture~\ref{conjecture:decomposition-conjecture} for the
linear-order perturbation $h_{ab}$ on spherically symmetric background
spacetimes~\cite{K.Nakamura-2021a}.
Then, we showed that above our formulation of a gauge-invariant
perturbation theory is applicable to perturbations on the
Schwarzschild spacetime including $l=0,1$ modes, and derived
the $l=0,1$ solutions to the linearized Einstein
equation~\cite{K.Nakamura-2021a}.

%*********************************************************************

From Eq.~(\ref{eq:nth-order-Einstein-eq}), the linearized Einstein
equation ${}^{(1)}\!G_{a}^{\;\;b}=8\pi {}^{(1)}\!T_{a}^{\;\;b}$  for
$h_{ab}=\ScrF_{ab}+{\pounds}_{Y}g_{ab}$ with the vacuum background
Einstein equation $G_{a}^{\;\;b}=8\pi T_{a}^{\;\;b}=0$ is given by
\begin{eqnarray}
  \label{eq:linear-Einstein-eq-gauge-inv}
  {}^{(1)}\!\ScrG_{a}^{\;\;b}\left[\ScrF\right]=8\pi {}^{(1)}\!\ScrT_{a}^{\;\;b},
\end{eqnarray}
and the linear-order continuity equations of the energy-momentum
tensor is given by
\begin{eqnarray}
  \nabla^{a}{}^{(1)}\!\ScrT_{a}^{\;\;b} = 0.
  \label{eq:divergence-barTab-linear-vac-back-u}
\end{eqnarray}
We decompose the components of the linear perturbation of
${}^{(1)}\!\ScrT_{ac}$ as
\begin{eqnarray}
  {}^{(1)}\!\ScrT_{ac}
  &\!\!\!\!\!\!\!\!
    =
    &\!\!\!\!\!\!\!\!
      \sum_{l,m}
      \tilde{T}_{AC}
      S_{\delta}
      (dx^{A})_{a} (dx^{C})_{c}
      +
      2
      r
      \sum_{l,m} \left\{
      \tilde{T}_{(e1)A} \hat{D}_{p}S_{\delta}
      +
      \tilde{T}_{(o1)A} \epsilon_{pq} \hat{D}^{q}S_{\delta}
      \right\}
      (dx^{A})_{(a} (dx^{p})_{c)}
      \nonumber\\
  &&
  \!\!\!\!\!\!\!\!\!\!\!\!\!\!\!\!
     +
     r^{2}
     \sum_{l,m} \left\{
     \tilde{T}_{(e0)} \frac{1}{2} \gamma_{pq} S_{\delta}
     +
     \tilde{T}_{(e2)} \left(
     \hat{D}_{p}\hat{D}_{q}
     -
     \frac{1}{2} \gamma_{pq} \hat{\Delta}
     \right) S_{\delta}
     +
     \tilde{T}_{(o2)} \epsilon_{s(p}\hat{D}_{q)}\hat{D}^{s}S_{\delta}
     \right\}
     (dx^{p})_{a} (dx^{q})_{c}
     .
     \label{eq:1st-pert-calTab-dd-decomp-2}
\end{eqnarray}

%*********************************************************************

Since we impose $\delta=0$ after deriving mode-by-mode perturbative
Einstein equations, we may choose $\tilde{T}_{(e2)}$ $=$
$\tilde{T}_{(o2)}$ $=$ $0$ for $l=0,1$ modes, and
$\tilde{T}_{(e1)A}=0=\tilde{T}_{(o1)A}$ for $l=0$ modes.
This choice and Eq.~(\ref{eq:divergence-barTab-linear-vac-back-u})
leads $\tilde{T}_{(e0)}=0$ for $l=0$ mode.
Then, we derived the $l=0,1$-mode solutions to
Eq.~(\ref{eq:linear-Einstein-eq-gauge-inv})~\cite{K.Nakamura-2021a}:

%*********************************************************************

For $l=1$ $m=0$ odd-mode perturbations, we derived
\begin{eqnarray}
  \label{eq:l=1-odd-mode-propagating-sol-ver2}
     2 {}^{(1)}\!\ScrF_{Ap}(dx^{A})_{(a}(dx^{p})_{b)}
  &\!\!\!\!\!\!\!\!=&\!\!\!\!\!\!\!
      \left(
      6M
      r^{2} \int dr
      \frac{1}{r^{4}} a_{1}(t,r)
      \right) \sin^{2}\theta (dt)_{(a}(d\phi)_{b)}
      +
      {\pounds}_{V_{(1,o1)}}g_{ab}
      ,
                      \\
  \label{eq:l=1-odd-mode-propagating-sol-ver2-Va-def}
  V_{(1,o1)a}
  &\!\!\!\!\!\!\!\!=&\!\!\!\!\!\!\!
  \left(\beta_{1}(t) + W_{(1,o)}(t,r)\right) r^{2} \sin^{2}\theta (d\phi)_{a}
  .
\end{eqnarray}
Here, $\beta_{1}(t)$ is an arbitrary function of $t$.
The function $a_{1}(t,r)$ is the solution to
Eq.~(\ref{eq:linear-Einstein-eq-gauge-inv}) given by
\begin{eqnarray}
  a_{1}(t,r)
  =
  - \frac{16 \pi}{3M} r^{3} f \int dt \tilde{T}_{(o1)r} + a_{10}
  =
  - \frac{16 \pi}{3M} \int dr r^{3} \frac{1}{f} \tilde{T}_{(o1)t} + a_{10}
  ,
  \label{eq:a1tr-sol}
\end{eqnarray}
where $a_{10}$ is the constant of integration which corresponds to the
Kerr parameter perturbation.
On the other hand, $rf \partial_{r}W_{(1,o)}$ of the variable $W_{(1,o)}$ in
Eq.~(\ref{eq:l=1-odd-mode-propagating-sol-ver2-Va-def}) is determined
by the evolution equation
\begin{eqnarray}
  \partial_{t}^{2}(r f \partial_{r}W_{(1,o)})
  -  f \partial_{r}( f \partial_{r}(r f \partial_{r}W_{(1,o)})
  + \frac{1}{r^{2}} f \left[ 3f-1 \right] (r f \partial_{r}W_{(1,o)})
  =
  16 \pi f^{2} \tilde{T}_{(o1)r}
  .
  \label{eq:odd-master-equation-Regge-Wheeler-l=1}
\end{eqnarray}

%*************************************************************

For the $l=0$ even-mode perturbation, we have
\begin{eqnarray}
  &&
     {}^{(1)}\!\ScrF_{ab}
     =
     \frac{2}{r}
     \left(
     M_{1}
     + 4 \pi \int dr \left[\frac{r^{2}}{f} \tilde{T}_{tt}\right]
     \right)
     \left(
     (dt)_{a}(dt)_{b}
     +
     \frac{1}{f^{2}}
     (dr)_{a}(dr)_{b}
     \right)
     \nonumber\\
  && \quad\quad\quad\quad
     +
     2 \left[
     4 \pi r \int dt \left(
     \frac{1}{f} \tilde{T}_{tt}
     + f \tilde{T}_{rr}
     \right)
     \right]
     (dt)_{(a}(dr)_{b)}
     +
     {\pounds}_{V_{(1,e0)}}g_{ab}
     ,
     \label{eq:l=0-final-sols}
  \\
  \label{eq:Va-second-choice-non-vac-sum}
  &&
     V_{(1,e0)a}
     :=
     \left(
     \frac{1}{4} f \Upsilon_{1} + \frac{1}{4} r f \partial_{r}\Upsilon_{1}
     + \gamma_{1}(r)
     \right)
     (dt)_{a}
     +
     \frac{1}{4f} r \partial_{t}\Upsilon_{1}
     (dr)_{a}
     ,
\end{eqnarray}
where $M_{1}$ is the linear-order Schwarzschild mass parameter
perturbation, $\gamma_{1}(r)$ is an arbitrary function of $r$.
The variable ${}^{(1)}\!\tilde{F}:=\partial_{t}\Upsilon_{1}$ in the
generator (\ref{eq:Va-second-choice-non-vac-sum}) satisfies the
following equation:
\begin{eqnarray}
  -  \frac{1}{f} \partial_{t}^{2}\tilde{F}
  + \partial_{r}( f \partial_{r}\tilde{F} )
  + \frac{1}{r^{2}} 3(1-f) \tilde{F}
  =
  - \frac{8}{r^{3}} m_{1}(t,r)
  + 16 \pi \left[
  -  \frac{1}{f} \tilde{T}_{tt}
  + f \tilde{T}_{rr}
  \right]
  ,
  \label{eq:even-mode-tildeF-eq-Phie-reduce}
\end{eqnarray}
where
\begin{eqnarray}
  m_{1}(t,r)
  =
  4 \pi \int dr \left[\frac{r^{2}}{f} \tilde{T}_{tt}\right]
  + M_{1}
  =
  4 \pi \int dt \left[ r^{2} f \tilde{T}_{rt} \right]
  + M_{1}
  ,
  \quad
  M_{1}\in\RF
  .
  \label{eq:Ein-non-vac-m1-sol}
\end{eqnarray}

%*************************************************************

For the $l=1$ $m=0$ even-mode perturbation, we have
\begin{eqnarray}
  &&
     {}^{(1)}\!\ScrF_{ab}
     =
     \frac{16\pi r^{2}}{1-f}
     \left\{
     - \frac{f^{2}}{3} \left[
     \frac{1+f}{2} \tilde{T}_{rr}
     + r f \partial_{r}\tilde{T}_{rr}
     -  \tilde{T}_{(e0)}
     -  4 \tilde{T}_{(e1)r}
     \right] (dt)_{a}(dt)_{b}
     \right.
     \nonumber\\
  && \quad\quad\quad\quad
     \left.
      +
      \left[
      (1-f) \tilde{T}_{tr}
      - \frac{2r}{3f} \partial_{t}\tilde{T}_{tt}
      \right]
     (dt)_{(a}(dr)_{b)}
     + \frac{1-3f}{2f^{2}} \left[
     \tilde{T}_{tt}
     -  \frac{2rf}{3(1-3f)} \partial_{r}\tilde{T}_{tt}
     \right]
     (dr)_{a}(dr)_{b}
     \right.
     \nonumber\\
  && \quad\quad\quad\quad
     \left.
     -  \frac{r^{2}\tilde{T}_{tt}}{3} \gamma_{ab}
     \right\} \cos\theta
     +
     {\pounds}_{V_{(1,e1)}}g_{ab}
      \label{eq:calFab-l=1-m=0-sol}
     ,
  \\
  &&
     V_{(1,e1)a}
     :=
     -  r \partial_{t}\Phi_{(e)} \cos\theta (dt)_{a}
     + \left( \Phi_{(e)} - r \partial_{r}\Phi_{(e)} \right) \cos\theta (dr)_{a}
     -  r \Phi_{(e)} \sin\theta (d\theta)_{a}
     ,
     \label{eq:generator-covariant-l=1-m=0-sum}
\end{eqnarray}
where $\Phi_{(e)}$ satisfies the following equation
\begin{eqnarray}
  &&
     \!\!\!\!\!\!\!\!\!\!\!\!\!\!
     -  \frac{1}{f} \partial_{t}^{2}\Phi_{(e)}
     + \partial_{r}\left[ f \partial_{r}\Phi_{(e)} \right]
     -
     \frac{1-f}{r^{2}} \Phi_{(e)}
     =
     16 \pi \frac{r}{3(1-f)} S_{(\Phi_{(e)})}
     ,
%      \label{eq:Zerilli-Moncrief-eq-final-l=1}
  \nonumber\\
  &&
     \!\!\!\!\!\!\!\!\!\!\!\!\!\!
     S_{(\Phi_{(e)})}
     :=
     \frac{3(1-3f)}{4f} \tilde{T}_{tt}
     -  \frac{1}{2} r \partial_{r}\tilde{T}_{tt}
     + \frac{1+f}{4} f \tilde{T}_{rr}
     + \frac{1}{2} f^{2} r \partial_{r}\tilde{T}_{rr}
     -  \frac{f}{2} \tilde{T}_{(e0)}
     -  2 f \tilde{T}_{(e1)r}
     .
     \label{eq:SPhie-def-explicit-l=1}
\end{eqnarray}

%*********************************************************************

%%%%%%%%%%%%%%%%%%%%%%%%%%%%%%%%%%%%%%%%%%%%%%%%%%%
%%%%%%%%%%%%%%%%%%%%%%%%%%%%%%%%%%%%%%%%%%%%%%%%%%%
%%%%%%%%%%%%%%%%%%%%%%%%%%%%%%%%%%%%%%%%%%%%%%%%%%%
%%%%%%%%%%%%%%%%%%%%%%%%%%%%%%%%%%%%%%%%%%%%%%%%%%%
%\section{Extension to the higher-order perturbations}
%\label{sec:higher-order_extension}
%%%%%%%%%%%%%%%%%%%%%%%%%%%%%%%%%%%%%%%%%%%%%%%%%%%
%%%%%%%%%%%%%%%%%%%%%%%%%%%%%%%%%%%%%%%%%%%%%%%%%%%
%%%%%%%%%%%%%%%%%%%%%%%%%%%%%%%%%%%%%%%%%%%%%%%%%%%
%%%%%%%%%%%%%%%%%%%%%%%%%%%%%%%%%%%%%%%%%%%%%%%%%%%

%*************************************************************

\paragraph{\bf 4. Extension to the higher-order perturbations} ----------
As shown in Sec.~2, the $n$-th order Einstein equation is given in
Eq.~(\ref{eq:nth-order-Einstein-eq}), which we rewrite as
\begin{eqnarray}
     {}^{(1)}\!\ScrG_{a}^{\;\;b}\left[{}^{(n)}\!\ScrF\right]
     =
     -
     {}^{({\rm NL})}\!\ScrG_{a}^{\;\;b}\left[
     \left\{ \left. {}^{(i)}\!\ScrF_{cd}\right| i<n \right\}
     \right]
     +
     8 \pi {}^{(n)}\!\ScrT_{a}^{\;\;b}
     =:
     8 \pi {}^{(n)}\!\TF_{a}^{\;\;b}
     .
     \label{eq:nth-einstein-gauge-inv-again}
\end{eqnarray}
Here, the left-hand side in
Eq.~(\ref{eq:nth-einstein-gauge-inv-again}) is the linear term of
${}^{(n)}\!\ScrF_{ab}$ and the first term in the right-hand side
is the non-linear term consists of the lower-order metric
perturbation ${}^{(i)}\!\ScrF_{ab}$ with $i<n$.
The right-hand side $8 \pi {}^{(n)}\!\TF_{a}^{\;\;b}$ of
Eq.~(\ref{eq:nth-einstein-gauge-inv-again}) is regarded an
effective energy-momentum tensor for the $n$th-order metric
perturbation ${}^{(n)}\!\ScrF_{ab}$.

%*************************************************************

The vacuum background condition $G_{a}^{\;\;b}=0$ implies the
identity $\nabla_{a}{}^{(1)}\!\ScrG_{b}^{\;\;a}\left[A\right]$ $=$
$0$ and Eq.~(\ref{eq:nth-einstein-gauge-inv-again}) implies
$\nabla^{a}{}^{(n)}\!\TF_{a}^{\;\;b} = 0$.
This equation gives consistency relations which should be confirmed.
Note that ${}^{(n)}\!\TF_{a}^{\;\;b}$ does not include
${}^{(n)}\ScrF_{ab}$, since the terms $-{}^{({\rm
    NL})}\!\ScrG_{a}^{\;\;b}\left[ \left\{\left. {}^{(i)}\!{\cal
        F}_{cd}\right| i<n \right\} \right]$ and
${}^{(n)}\ScrT_{a}^{\;\;b}$ in
Eq.~(\ref{eq:nth-einstein-gauge-inv-again}) don't include
${}^{(n)}\ScrF_{ab}$ due to the vacuum background condition.
This situation is same as that when we solved the linear equations
(\ref{eq:linear-Einstein-eq-gauge-inv})--(\ref{eq:divergence-barTab-linear-vac-back-u}).
Furthermore, we decompose ${}^{(n)}\!\TF_{ab}$ as
\begin{eqnarray}
  &&
     \!\!\!\!\!\!\!\!\!\!\!\!\!\!\!\!\!
     {}^{(1)}\!\TF_{ab}
     =:
     \sum_{l,m}
     \tilde{\TF}_{AB}
     S_{\delta}
     (dx^{A})_{a} (dx^{B})_{b}
     +
     2
     r
     \sum_{l,m} \left\{
     \tilde{\TF}_{(e1)A} \hat{D}_{p}S_{\delta}
     +
     \tilde{\TF}_{(o1)A} \epsilon_{pq} \hat{D}^{q}S_{\delta}
     \right\}
     (dx^{A})_{(a} (dx^{p})_{b)}
     \nonumber\\
  &&
     +
     r^{2}
     \sum_{l,m} \left\{
     \tilde{\TF}_{(e0)} \frac{1}{2} \gamma_{pq} S_{\delta}
     +
     \tilde{\TF}_{(e2)} \left(
     \hat{D}_{p}\hat{D}_{q}
     -
     \frac{1}{2} \gamma_{pq} \hat{\Delta}
     \right) S_{\delta}
     +
     \tilde{\TF}_{(o2)} \epsilon_{s(p}\hat{D}_{q)}\hat{D}^{s}S_{\delta}
     \right\}
     (dx^{p})_{a} (dx^{q})_{b}
     .
     \label{eq:1st-pert-TFab-dd-decomp}
\end{eqnarray}
Then, the replacements
$\tilde{T}_{AB}\rightarrow \tilde{\TF}_{AB}$,
$\tilde{T}_{(e1)A}\rightarrow \tilde{\TF}_{(e1)A}$,
$\tilde{T}_{(o1)A}\rightarrow \tilde{\TF}_{(o1)A}$,
$\tilde{T}_{(e0)}\rightarrow \tilde{\TF}_{(e0)}$,
$\tilde{T}_{(e2)}\rightarrow \tilde{\TF}_{(e2)}$,
$\tilde{T}_{(o2)}\rightarrow \tilde{\TF}_{(o2)}$
in the solutions
(\ref{eq:l=1-odd-mode-propagating-sol-ver2})--(\ref{eq:SPhie-def-explicit-l=1})
yield the solutions to Eq.~(\ref{eq:nth-einstein-gauge-inv-again}).

%*************************************************************

%%%%%%%%%%%%%%%%%%%%%%%%%%%%%%%%%%%%%%%%%%%%%%%%%%%%%%
%%%%%%%%%%%%%%%%%%%%%%%%%%%%%%%%%%%%%%%%%%%%%%%%%%%%%%
%%%%%%%%%%%%%%%%%%%%%%%%%%%%%%%%%%%%%%%%%%%%%%%%%%%%%%
%\section{Summary}
%\label{sec:summary_and_discussion}
%%%%%%%%%%%%%%%%%%%%%%%%%%%%%%%%%%%%%%%%%%%%%%%%%%%%%%
%%%%%%%%%%%%%%%%%%%%%%%%%%%%%%%%%%%%%%%%%%%%%%%%%%%%%%
%%%%%%%%%%%%%%%%%%%%%%%%%%%%%%%%%%%%%%%%%%%%%%%%%%%%%%

%****************************************************************

\paragraph{\bf 5. Summary} ----------
We proposed a gauge-invariant treatment of the $l=0,1$-mode
perturbations on the Schwarzschild background spacetime as the
Proposal~\ref{proposal:harmonic-extension}.
Following this proposal, we derived the $l=0,1$-mode solutions to the
Einstein equations with the general linear perturbations of the
energy-momentum tensor in the gauge-invariant manner.

%****************************************************************

The derived solution in the $l=1$ odd mode actually realizes the
linearized Kerr solution in the vacuum case.
Furthermore, we also derived the $l=0,1$ even-mode solutions to the
Einstein equations.
In the vacuum case, in which all components of ${}^{(1)}\!\ScrT_{ab}$
vanish, the $l=0$ even-mode solution realizes the only the additional
mass parameter perturbation of the Schwarzschild spacetime.
These results are the realization of the linearized gauge-invariant
version of uniqueness theorem of Kerr black hole and these solutions
are physically reasonable.
Owing to this realization, we may say that our proposal is also
physically reasonable.
Details of our discussions are given in Ref.~\cite{K.Nakamura-2010}.

%****************************************************************

The fact that we confirmed
Conjecture~\ref{conjecture:decomposition-conjecture} for the
linear-metric perturbations in the Schwarzschild background case
including the $l=0,1$ modes implies that the extension to any-order
perturbations through our gauge-invariant
formulation~\cite{K.Nakamura-2014} was possible, at least, in the case
of the Schwarzschild background case.
Thus, we can develop a higher-order gauge-invariant perturbation
theory on the Schwarzschild background
spacetime~\cite{K.Nakamura-2021b}.

%*******************************************************************

We leave the development for specific astrophysical situations such
as gravitational-wave astronomy through our formulation as future
works.

%*******************************************************************

%%%%%%%%%%%%%%%%%%%%%%%%%%%%%%%%%%%%%%%%%%%%%%%%%%%%%%
%%%%%%%%%%%%%%%%%%%%%%%%%%%%%%%%%%%%%%%%%%%%%%%%%%%%%%
%%%%%%%%%%%%%%%%%%%%%%%%%%%%%%%%%%%%%%%%%%%%%%%%%%%%%%
%\section*{References}
%%%%%%%%%%%%%%%%%%%%%%%%%%%%%%%%%%%%%%%%%%%%%%%%%%%%%%
%%%%%%%%%%%%%%%%%%%%%%%%%%%%%%%%%%%%%%%%%%%%%%%%%%%%%%
%%%%%%%%%%%%%%%%%%%%%%%%%%%%%%%%%%%%%%%%%%%%%%%%%%%%%%

%% Full authors list (ONLY FOR COLLABORATIONS)
%\clearpage
%\section*{Full Authors List: \Coll\ Collaboration}
%
%\noindent \textbf{Note comment afterwards:} Collaborations have the possibility to provide an authors list in xml format which will be used while generating the DOI entries making the full authors list searchable in databases like Inspire HEP. \\
%
%\scriptsize
%\noindent
%first.author$^1$, 
%second.author$^2$, 
%third.author$^3$ % .... more names
%and 
%last.author$^{n}$ \\
%
%\noindent
%$^1$first.affiliation.
%$^2$second.affiliation. % .... more affiliation
%$^{m}$last.affiliation.

\end{document}